\documentclass[twoside]{article}
\usepackage{fleqn,espcrc2}

\usepackage{graphicx}

\newcommand{\AmS}{{\protect\the\textfont2
  A\kern-.1667em\lower.5ex\hbox{M}\kern-.125emS}}

\title{Josephson Plasma Mode in Fields Parallel to Layers of
	Bi$_2$Sr$_2$CaCu$_2$O$_{8+\delta}$}

\author{Itsuhiro Kakeya
		\address{Institute of Materials Science, 
		University of Tsukuba, \\ 
        1-1-1 Ten-nodai, Tsukuba, Ibaraki 305-8573, Japan}$^\mathrm{b}$,
        Tomoyuki Wada~$^\mathrm{a}$, Ryo Nakamura~$^\mathrm{a}$,
        and Kazuo Kadowaki~$^\mathrm{a}\!\!$
        \address{CREST, Japan Science and Technology Cooperation}}
        
\begin{document}

\begin{abstract}
Josephson plasma resonance measurements under magnetic fields parallel to the CuO$_2$ layers as functions of magnetic field, temperature, and microwave frequency have been performed in Bi$_2$Sr$_2$CaCu$_2$O$_{8+\delta}$ single crystals with doping range being from optimal to under-doped side.
The feature of the resonance is quite unique and cannot be explained by the conventional understandings of the Josephson plasma for $H \parallel c$, that requires a new theory including coupling effect between Josephson vortex lattice and Josephson plasma.
\vspace{1pc}
\end{abstract}

\maketitle

Recent studies of Josephson plasma resonance (JPR) in fields perpendicular to CuO$_2$ $ab$ plane of Bi$_2$Sr$_2$CaCu$_2$O$_{8+\delta}$ have revealed a plenty of intriguing phenomena in vortex states of high-$T_c$ superconductors in these several years.
Though the features of JPR in the vortex states in $H \parallel c$ have almost been comprehended, JPR in fields parallel to the $ab$ plane is quite controversial both theoretically and experimentally~\cite{JPR2}.
As is known, Josephson vortices are created inside block layers in this orientation, and their oscillation mode is expected to couple strongly with the Josephson plasma mode~\cite{Sonin}.
In the present work, we have performed systematic JPR measurements in $H \parallel ab$ at various microwave frequencies in optimum and under-doped crystals.
The results cannot be interpreted by the theory of JPR used for $H \parallel c$, and imply a coupled mode with the oscillation of the Josephson vortex lattice.

We have used three under-doped (U1, U2, and U3) and a optimally-doped (OP) Bi$_2$Sr$_2$CaCu$_2$O$_{8+\delta}$ single crystals, which are grown by the modified TSFZ method and annealed in proper atmospheres.
The superconducting transition temperatures ($T_c$) obtained by magnetization measurements are 70, 72, 78, and 91 K for U1, U2, U3, OP, respectively.
JPR measurements were performed at five frequencies of 18.9, 25.5, 34.3, 44.2, and 52.3 GHz by a reflective type of bridge balance circuit with $TE_{102}$ mode rectangular cavity resonators.
Samples were placed inside the cavity so that the microwave electric field was parallel to the $c$ axis, in which configuration only the longitudinal Josephson plasma mode is excited~\cite{Kake}.
The resonance curves were obtained by sweeping temperature at a fixed magnetic field.
The external field was applied by a split-pair of superconducting magnet and the direction was varied with rotating the microwave setup including the cavity with respect to the magnet by a high-precision goniometer.
The direction of the $ab$ plane was determined by looking at the symmetry of angular dependence of JPR lines.

\begin{figure}[tb]
\begin{center}
\includegraphics[width=\linewidth]{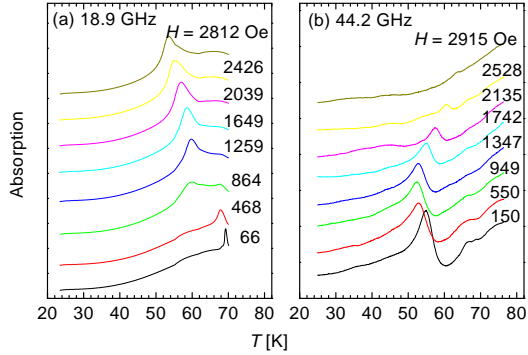}
\end{center}
\vspace{-1cm}
\caption{Resonance curves at 19.8 (a) and 44.2 (b) GHz in U1 in $H \parallel ab$.}
\label{fig:line}
\end{figure}

Figure \ref{fig:line} represents resonance curves obtained at 18.9 and 44.2 GHz in sample U1.
At 18.9 GHz (Fig. 1(a)), two kinds of resonance lines are observed: one is observed in higher temperature and lower field region and the other is observed in lower temperature and higher field region. 
We call them HTB (higher temperature branch) and LTB (lower temperature branch) for the former and the latter, respectively.
HTB is observed at almost zero field as a quite sharp and symmetric line at 68 K, is broadened as the parallel field increases, and finally disappears above 1000 Oe. 
The resonance temperature slightly shifts lower temperature at low fields, and changes little above 600 Oe.
LTB begins to appear at 500 Oe with a broad line from lower temperature and shift to higher temperature as the external field increases. 
Above 1500 Oe, the line becomes clearer and the resonance temperature tends to decrease.
As the microwave frequency increases, both branches shift to lower temperatures, and only HTB with relatively broader line-shape is observed at 44.2 GHz as shown in Fig. 1(b).
The resonance temperature shows significant non-monotonic field dependence: as the field increases, the resonance temperature once decreases then increases above 1000 Oe, and finally exceeds the resonance temperature at zero-field.
Such splitting of the resonance mode is observed also in U2 and U3 at 34.2 GHz, and in OP at 44.2 GHz, although the features depend upon anisotropies of crystals and the microwave frequencies.

\begin{figure}[tb]
\includegraphics[width=\linewidth]{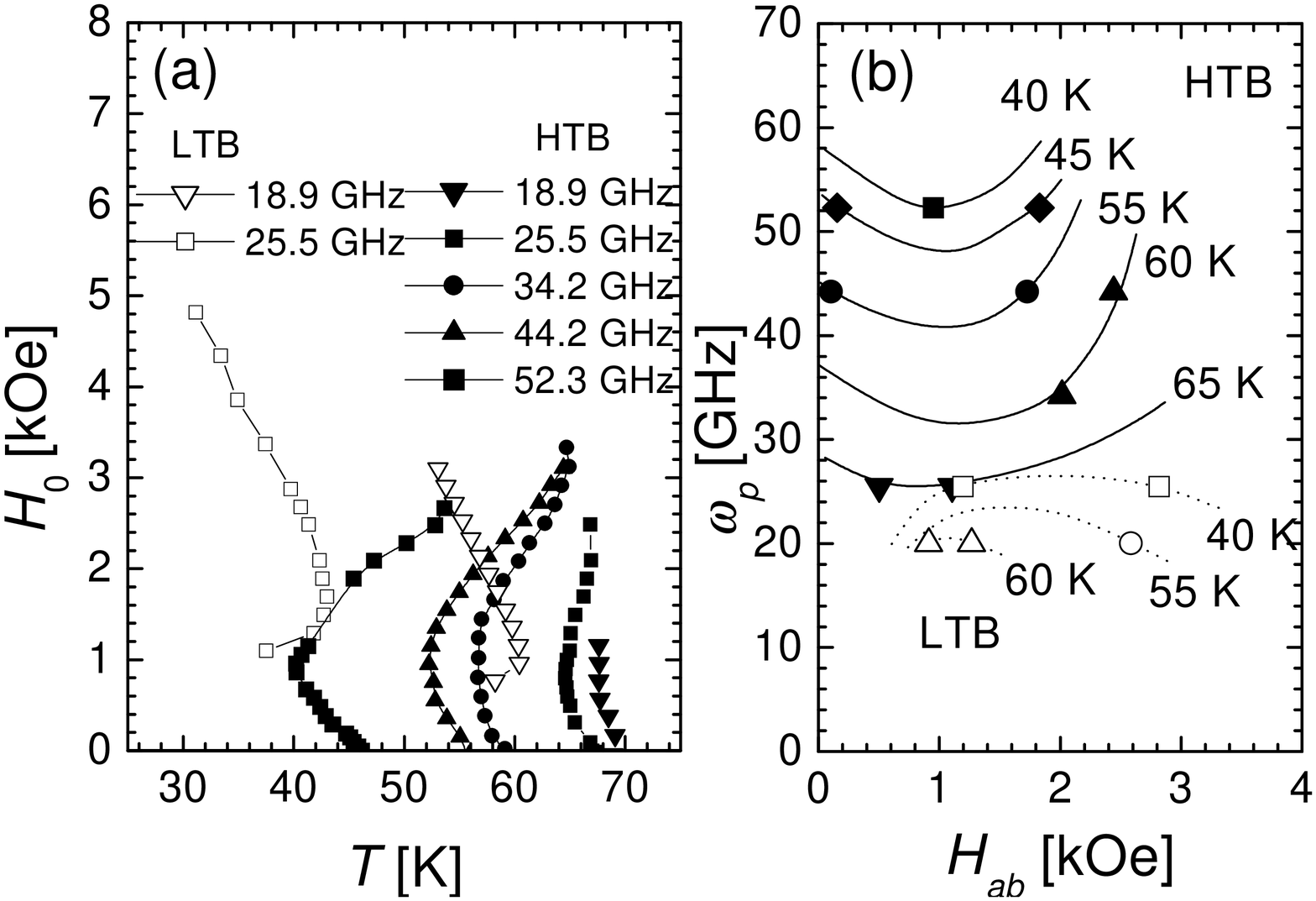}
\vspace{-1cm}
\caption{(a) Plots of resonance peaks for HTB (solid symboles) and LTB (open symboles). (b) Plasma frequency $\omega_p$ as a functions of in-plane field $H_{ab}$. 
Solid and dotted lines are for the eyes.}
\label{}
\end{figure}

HTB and LTB observed in sample U1 are plotted for various microwave frequencies in Fig. 2(a).
In order to clarify the in-plane field contribution for the plasma mode, plasma frequency $\omega_p$ of U1 is plotted as a function of the external field $H_{ab}$ for various temperatures in Fig. 2(b).
At all temperatures, $\omega_p$ of HTB make minimum around 1000 Oe and has finite value at $H_{ab}=0$ corresponding to the Josephson plasma frequency at zero field. From this reason, HTB is interpreted as the Josephson plasma mode strongly modified with Josephson vortices.
On the other hand, LTB has a maximum at finite fields depending upon temperatures. It seems to be $\omega_p \to 0$ at $H_{ab} \to 0$.
Such a distinct difference of the field dependence between two modes indicates that LTB has a different origin from HTB.
The behavior of LTB at $H_{ab} \to 0$ implies that LTB is strongly related to the oscillation mode of Josephson vortex lattice.
To determine the origin of LTB, measurements in lower frequency region are required.

In summary, we have observed two split Josephson plasma modes in magnetic field orientations being parallel to the $ab$ plane.
From the in-plane field dependence of the plasma frequency, we argue that these two modes come from different origins: one named HTB originates from the intrinsic Josephson plasma mode for $H \parallel ab$, and the other named LTB may come from a coupled oscillation mode with the Josephson vortex lattice.

\end{document}